\documentclass[twocolumn,times,astrosymb]{aastex631}

\usepackage{CJK}
\usepackage{xspace}
\usepackage{amssymb}
\usepackage{xcolor}

\newcommand{\Gaia}{\textit{Gaia}\xspace}
\newcommand{\TESS}{\textit{TESS}\xspace}
\newcommand{\Porb}{\ensuremath{P_{\rm orb}}}
\newcommand{\DRV}{\ensuremath{\Delta V_{\rm r}}}
\newcommand{\semiDRV}{\ensuremath{\Delta V_{\rm r} / 2}}
\newcommand{\massfunc}{\ensuremath{f(M_2)}}
\newcommand{\Msun}{\ensuremath{M_\odot}}
\newcommand{\Rsun}{\ensuremath{R_\odot}}
\newcommand{\Vr}{\ensuremath{V_{\rm r}}}
\newcommand{\Teff}{\ensuremath{T_{\rm eff}}}
\newcommand{\kms}{km s$^{-1}$\xspace}

\graphicspath{{./figures/}}

\shorttitle{Searching for compact objects with \Gaia DR3}
\shortauthors{Fu et al.}

\begin{document}
\begin{CJK*}{UTF8}{gbsn}

  \title{Searching for compact objects in binaries with \Gaia DR3}

  \correspondingauthor{Wei-Min Gu, Zhi-Xiang Zhang, Tuan Yi}
  \email{guwm@xmu.edu.cn \\
    zhangzx@xmu.edu.cn \\
    yit@xmu.edu.cn}

  \author[0000-0003-2896-7750]{Jin-Bo Fu}
  \affiliation{Department of Astronomy, Xiamen University, Xiamen, Fujian 361005, China}

  \author[0000-0003-3137-1851]{Wei-Min Gu}
  \affiliation{Department of Astronomy, Xiamen University, Xiamen, Fujian 361005, China}

  \author[0000-0002-2419-6875]{Zhi-Xiang Zhang}
  \affiliation{Department of Astronomy, Xiamen University, Xiamen, Fujian 361005, China}

  \author[0000-0002-5839-6744]{Tuan Yi}
  \affiliation{Department of Astronomy, Xiamen University, Xiamen, Fujian 361005, China}

  \author[0000-0002-7135-6632]{Sen-Yu Qi}
  \affiliation{Department of Astronomy, Xiamen University, Xiamen, Fujian 361005, China}

  \author[0000-0002-5630-7859]{Ling-Lin Zheng}
  \affiliation{Department of Astronomy, Xiamen University, Xiamen, Fujian 361005, China}

  \author[0000-0002-2874-2706]{Jifeng Liu}
  \affiliation{National Astronomical Observatories, Chinese Academy of Sciences, Beijing 100101, China}
  \affiliation{College of Astronomy and Space Science, University of Chinese Academy of Sciences, Beijing 100049, China}

  \begin{abstract}

    We search for compact objects in binaries based on \Gaia~DR3. A sample of ten targets is derived under the conditions: radial velocity variable, low temperature ($\Teff < 6000$ K), high mass function ($\massfunc > 1 \Msun$), and ellipsoidal-like light curves. Two targets have LAMOST spectroscopic observations, one of which is a double-lined spectroscopic binary. The observational data of seven targets are not self-consistent, since their photometric periods are even shorter than the theoretical minimum orbital periods calculated by the stellar parameters from \Gaia DR3. After excluding these seven inconsistent targets and another target contaminated by a near-bright star, the remaining two targets may contain compact objects worth follow-up observations. This work may serve as an example to demonstrate the feasibility of searching for compact objects in the massive \Gaia data.

  \end{abstract}


  \section{Introduction} \label{sec:intro}

  Dynamical searching and identification of binary systems by spectroscopic
  surveys has made progress in recent years, especially for quiescent
  (e.g., without accretion) binary systems that contain compact objects
  \citep[black holes, neutron stars, and white dwarfs, e.g.,][]{2014Natur.505..378C,
    2019Sci...366..637T,2019Natur.575..618L,2020A&A...637L...3R,2021MNRAS.504.2577J,
    2021MNRAS.501.1677H,Saracino2022,2022ApJ...933L..23L,2022arXiv220707675S,2022arXiv220611270M}.
  Several such compact object candidates have been proposed from spectroscopic
  surveys \citep[e.g.][]{2016MNRAS.458.3808R,
    2017MNRAS.470.1442C,2017MNRAS.472.4193R,2018MNRAS.477.4641R,2019ApJ...872L..20G,
    2019AJ....158..179Z,2022SCPMA..6529711M,2022ApJ...933..193Z}.
  Besides, X-ray surveys are vital for searching for compact objects within
  accreting binaries \citep{2006ARA&A..44...49R,2006csxs.book..157M}, and gravitational
  wave observations are successful in detecting compact objects from mergers \citep{2016PhRvX...6d1015A}.
  These three approaches provide plentiful candidates that can be used
  to study the mass distributions and characterize the populations of compact objects.

  The samples of compact object candidates can be potentially enlarged greatly \citep[e.g.][]{2022arXiv220606032G}
  thanks to the \Gaia Data Release 3 \citep[\Gaia DR3;][]{Recio2022}, which provides astrometry,
  photometry, and spectroscopy for stars with an unprecedented scale. For instance,
  astrometry analyses are expected to uncover about 190 compact binaries from \Gaia DR3
  \citep{2022A&A...658A.129J}. Recently, a promising candidate \Gaia BH1 has been proposed
  utilizing the astrometric solution from \Gaia DR3 \citep{2022arXiv220906833E}.
  Furthermore, the radial velocity variations ($\DRV$) of 6,981,248 targets taken
  with the RVS instrument are firstly among the available data products \citep{Recio2022},
  which can facilitate the searching for compact objects from a dynamical perspective.
  The wavelength coverage of the $\Vr$ instrument is 846-870 nm, with a typical spectral resolution of about 11,500
  \citep{2022arXiv220606143D,2022arXiv220606205M}. \Gaia DR3 also provides a variety of labels for each object
  \citep{2022arXiv220605439H}, like single-lined spectroscopic binary (SB1) and double-lined spectroscopic binary (SB2) classifications,
  temperatures and radii \citep{2022gdr3.reptE..20H}, which contribute to better characterizing
  the parameters of the stellar systems.

  A proper method can facilitate the filtering of interesting targets from
  \Gaia's massive data products. The minimum orbital period ($\Porb^{\rm min}$)
  of a binary system can be calculated using the mass ($M_1$) and radius ($R_1$)
  of its visible component \citep{2019ApJ...872L..20G,2019AJ....158..179Z}, and the minimum mass function
  $\massfunc$ of its invisible component can be calculated using the $\DRV$
  of its hosting system. Therefore, by requiring a lower limit on the mass
  function, we can select potential hidden compact objects in binary systems
  that are not visible optically.

  The above approach relies only on the data provided by \Gaia. Nevertheless,
  more data would increase the credibility of the selected samples. For
  instance, multi-band spectral energy distributions (SEDs) and light curves contribute to judging whether
  a binary system is composed of two visible stars \citep{elbadry2022}.
  The spectroscopic surveys (e.g. the Large Sky Area Multi-Object Fiber
  Spectroscopic Telescope; LAMOST; see \citealt{cui2012,zhao2012}) can help
  to determine whether the target is a binary system \citep{Merle2017,lichunqian2021},
  and whether the radial velocity variation is credible.

  Based on the above methodology, we first select a raw sample with high mass function
  from \Gaia DR3, then utilize the high-precision light curves from the Transiting Exoplanet Survey Satellite \citep[\TESS,][]{2015JATIS...1a4003R}
  to find ellipsoidal variables \citep{1985ApJ...295..143M,1993ApJ...419..344M,2021MNRAS.501.2822G,2021MNRAS.507..104R},
  and fitting the broad-band SEDs to validate the radii and temperatures provided by \Gaia.
  Further validation is done by using spectroscopy from LAMOST
  to exclude binary stars and derive the candidates.

  The rest of this paper is organized as follows.
  Observations and data reduction are described in Section \ref{sec:data}.
  The selected targets are analyzed in detail in Section \ref{sec:discuss},
  followed by our main conclusions in Section \ref{sec:conclusion}.

  \section{Data Selection and Processing} \label{sec:data}

  \subsection{\Gaia}

  We begin by filtering the \Gaia DR3 database\footnote{\url{https://gea.esac.esa.int/archive/}}
  with the following criteria to identify the velocity variable stars and obtain their
  radial velocity variations $\DRV$ (\texttt{rv\_amplitude\_robust}):

  \begin{itemize}
    \item \texttt{rv\_nb\_transits} $\geqslant 10$,
    \item \texttt{rv\_template\_teff} $\leqslant 8000$ K,
    \item \texttt{rv\_chisq\_pvalue} $\leqslant 0.01$,
    \item \texttt{rv\_renormalised\_gof} $> 4$, and
    \item \texttt{teff\_gspphot} $< 6000$ K,
  \end{itemize}

  \noindent where the first four criteria are suggested by \citet{2022arXiv220605902K}, and the last one
  is to ensure the exclusion of hot stars (defined as $\Teff > 6000$ K), which may suffer more considerable RV
  uncertainties \citep{2022arXiv220605486B}. The above RV cuts reduced the number of candidates to 328,621.
  We additionally applied astrometric and photometric quality cuts to exclude targets with large
  distance measurement errors and potentially inaccurate photometry, which further shrinks the
  number of targets to 317,278:

  \begin{itemize}
    \item \texttt{parallax\_over\_error} $> 5$,
    \item \texttt{phot\_bp\_rp\_excess\_factor} $< 3$.
  \end{itemize}

  A more detailed description of these properties can be found in \citet{2022gdr3.reptE..20H}.
  We require that the masses and radii provided by \Gaia are not null,
  and use them to obtain the theoretical minimum orbital period of
  the remaining 284,607 binary systems \citep{2019AJ....158..179Z}:

  \begin{equation} \label{eq:1}
    P_{\rm orb}^{\rm min} = 2 \pi \left[ \frac{(R_1 / 0.462)^3}{G M_1} \right]^{1/2},
  \end{equation}

  \noindent where $M_1$, $R_1$ is the mass and radius of the optically visible star.
  The mass function of the invisible star can then be expressed as \citep{2006ARA&A..44...49R}:

  \begin{equation} \label{eq:2}
    f(M_2) = \frac{M_2^3 \sin^3(i)}{(M_1 + M_2)^2} = \frac{K_1^3 \Porb}{2\pi G},
  \end{equation}

  \noindent where $K_1 \geqslant \semiDRV$ is the semi-amplitude of the radial velocity,
  and $\Porb \geqslant P_{\rm orb}^{\rm min}$ is the orbital period of the target.
  Finally, to select those potential compact objects, we require
  \begin{itemize}
    \item $\massfunc > 1 \ \Msun$.
  \end{itemize}
  The above steps yield 65 targets from the \Gaia DR3 database.

\begin{figure*}
    \includegraphics[width=0.99\textwidth]{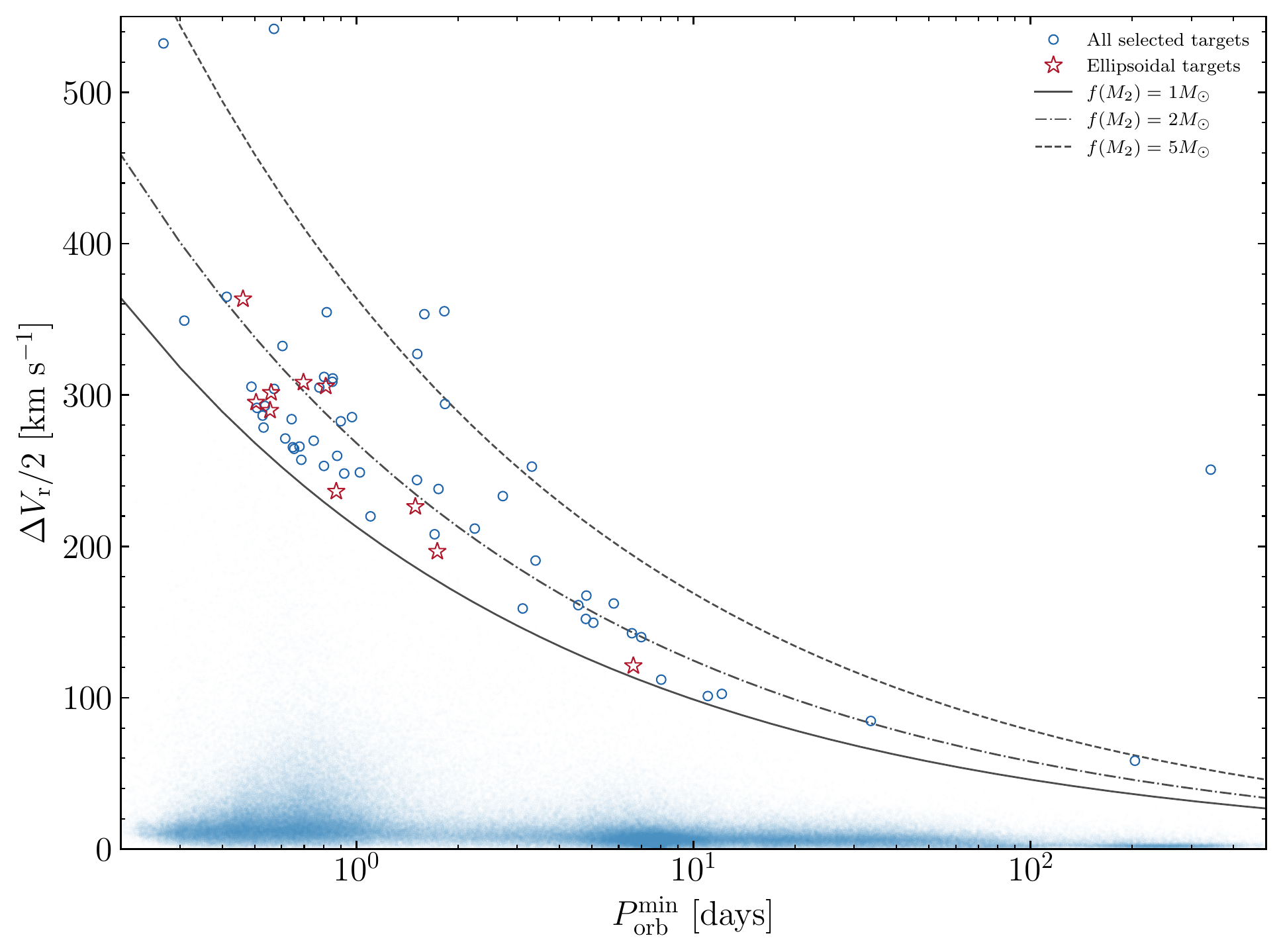}
    \caption{A sample of 65 targets selected from \Gaia~DR3. The ten red
      pentagrams indicate the targets presented in Table~\ref{table:1}.
      The solid, dot-dashed, and dashed lines each corresponds to a fixed
      value of the mass function. To demonstrate the distribution of the mass functions,
      we show all the 278,536 targets that satisfy the criteria described in Section \ref{sec:data}
      for comparison, represented by blue points (blurred for reducing the size of the figure).}
    \label{fig:All_Selected_Samples_with_Theoretical_Period}
  \end{figure*}

  Figure \ref{fig:All_Selected_Samples_with_Theoretical_Period} shows the theoretical minimum orbital period and semi-amplitude of velocity
  variations of the 65 selected targets. Two targets (Gaia ID: 5200508962219751424 and
  2066328011855804544) have theoretical minimum orbital periods $\Porb^{\rm min} > 100$ days.
  The long periods are due to their large radii of $178 \ \Rsun$ and $125 \ \Rsun$ measured by
  Gaia (see supplement Table \ref{table:2}). Their radii are much larger than
  other selected targets ($\sim 8.5 \ \Rsun$ on average), leading to
  longer theoretical minimum orbital periods (Equation \ref{eq:1}).
  These two objects are excluded from our final samples due to their
  \TESS light curves (see Section \ref{sec:TESS}).

  \subsection{\TESS photometry} \label{sec:TESS}

\begin{figure*}
    \includegraphics[width=0.99\textwidth]{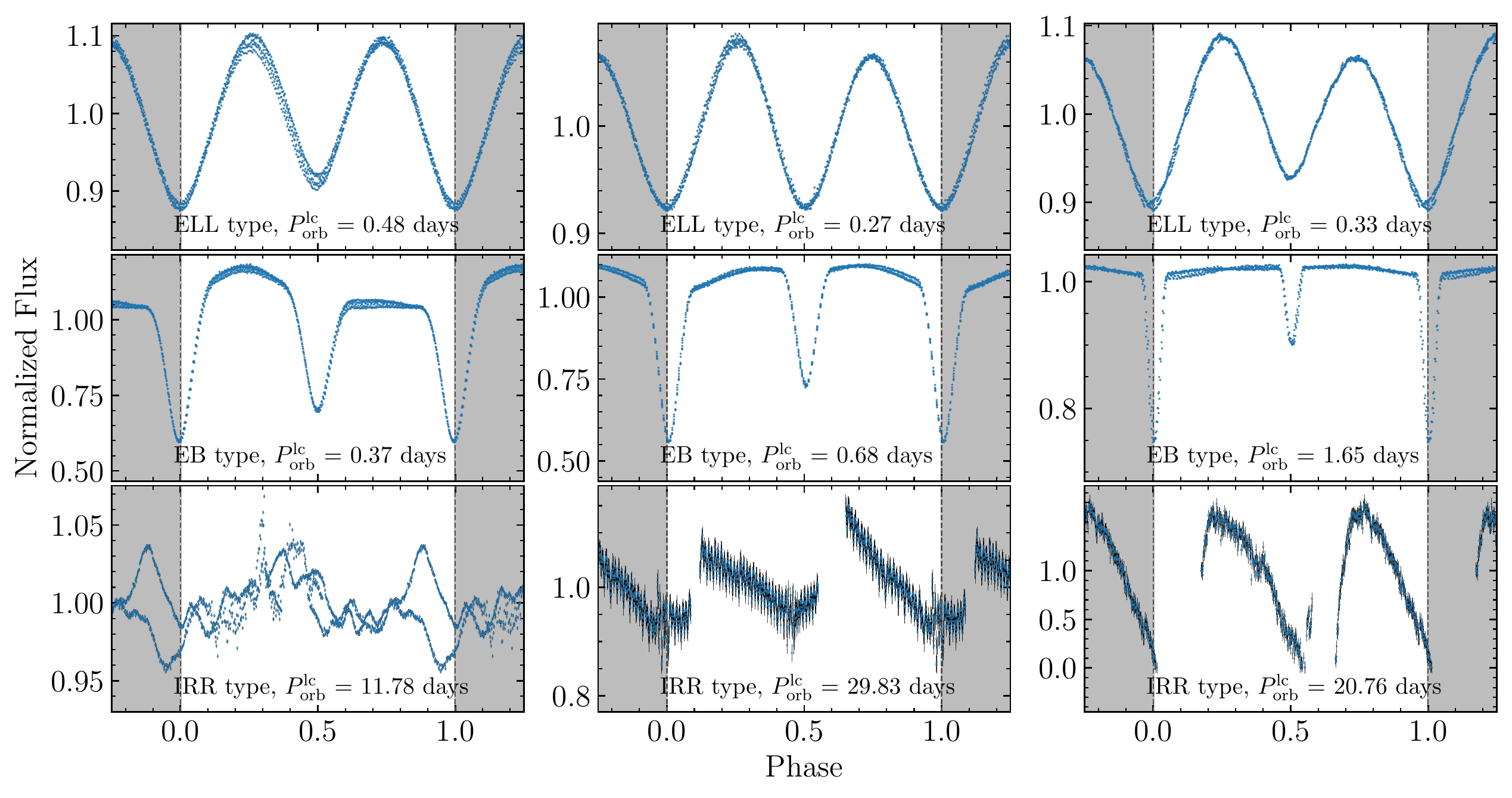}
    \caption{Representative \TESS light curve samples for ellipsoidal variables (ELLs, upper panels),
      eclipsing binaries (EBs, middle panels), and irregular variables (IRRs, bottom panels).
      The three ellipsoidal examples correspond to J1631, J1227, and J0056 listed in
      Table \ref{table:1}, from left to right. The three eclipsing binary samples are \Gaia
      714884906850837120, 2922949687038297088, and 4853820950132560896. The three irregular variables
      are \Gaia 2066328011855804544, 2054348862739979648, and 346272671566816512. The properties of eclipsing
      binaries and irregular variables can be found in Table \ref{table:2}.
      \label{fig:Lightcurve_Examples}
    }
  \end{figure*}

\begin{figure*}
    \includegraphics[width=0.99\textwidth]{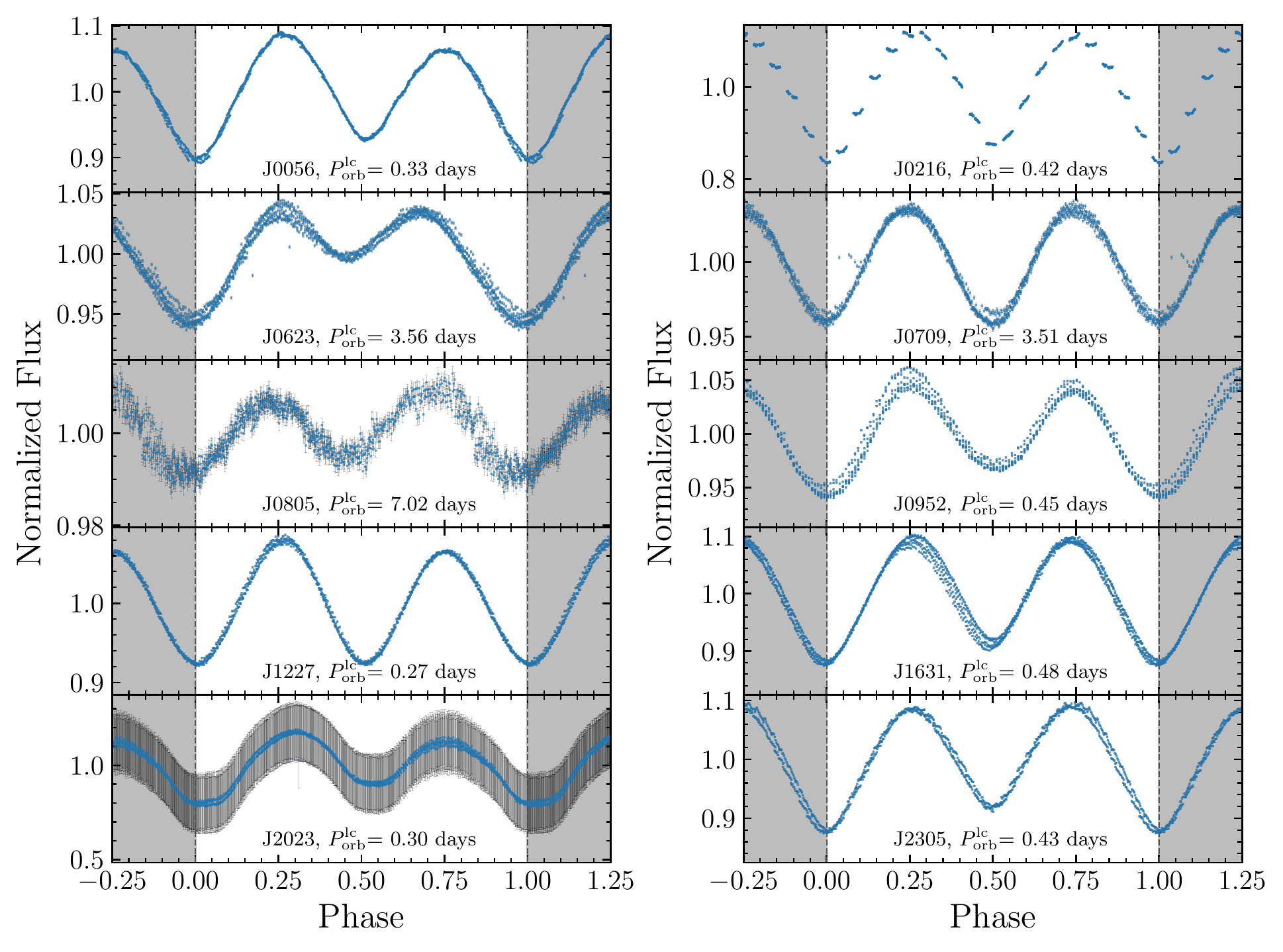}
    \caption{The phase-folded \TESS light curves of the ten targets in Table \ref{table:1}.
      The primary minimum of each light curve has been shifted to phase 0.
      Light curves beyond phase 0 - 1 (shaded with gray background) are repeated for clarity.}
    \label{fig:Lightcurves_for_Ten_Selected_Targets}
  \end{figure*}

  Using \texttt{lightkurve}\footnote{\url{https://docs.lightkurve.org}}, we retrieve
  the 65 targets listed in Table \ref{table:2} and download available
  photometric data for 55 of them from \TESS.
  We obtain the photometric period for each target using the Lomb-Scargle periodogram \citep{1989ApJ...338..277P},
  which is then used to phase-fold the light curve. For targets with multiple \TESS sectors observed,
  the sector with the longest observation time span is used. We reject targets with a folded photometric
  period $P_{\rm orb}^{\rm lc} > 15$ days, since each \TESS sector has at most 27 days of time coverage
  \citep{2015JATIS...1a4003R}. We visually inspect and manually classify the phase-folded light curves into:
  (1) ellipsoidal variables (tidally distorted stars which may contain a hidden compact object),
  (2) eclipsing binaries (binary stars show occultations periodically), and
  (3) irregular variables (such as eruptive stars, spotted rotational variables, variability of unknown origin,
  and variability of non-variables due to systematic effects of \TESS).
  Figure \ref{fig:Lightcurve_Examples} shows three cases for each variable class as examples.
  We found and discarded 29 eclipsing binaries with 16 irregular variables.
  The remaining ten sources have ellipsoidal-like light curves (Figure \ref{fig:Lightcurves_for_Ten_Selected_Targets}) and are chosen as the final sample.
  The photometric periods of these ten targets, and their mass functions calculated
  from the photometric periods, are listed in Table \ref{table:1}.
  Note that the lack of ellipsoidal variability cannot rule out potential massive unseen companions
  since the ellipsoidal modulation is sensitive to the separation of the two stars.
  For instance, long period (e.g. $\gtrsim$ 10 days) detached systems may not exhibit ellipsoidal variabilities
  where a main sequence companion could be subject to negligible tidal distortion from a compact companion.
  In this paper, we focus on ellipsoidal variables for simplicity.

  \subsection{SED fitting}

  We measure the surface temperatures and radii of the visible stars
  of our selected objects by SED fitting.
  We use the SED fitting to derive robust stellar parameters
  and therefore to validate that of provided by \Gaia DR3.
  Furthermore, we check the goodness-of-fit to evaluate whether a single-star SED
  model can fit well the data, and thus to verify the possible
  contaminations from binary systems that contain two stellar components.
  However, it should be noted that a good SED fitting by a single stellar component
  does not mean that the SED only has one component. For example, a binary
  system composed of two stars with similar temperatures can also be
  well-fitted with a single component.

  The fitting is carried out with the help of the \texttt{astroARIADNE}\footnote{\url{https://github.com/jvines/astroARIADNE}},
  which can automatically download the broadband SED of a given target from
  UV to IR band and fit it using a set of alternative atmospheric models.
  The effective temperature ($\Teff$), radius ($R_1$), distance, log surface
  gravity ($\log g$), metallicity ($\rm{[Fe/H]}$), and line-of-sight extinction ($A_V$) are
  six fitted parameters. Where the distance and the $R_1$ are
  degenerate parameters, we use the distance measured by \Gaia DR3 as
  prior to eliminate the degeneracy. The G-band extinctions from \Gaia
  are used as a mean of Gaussian prior for each target, since some of them are located near the
  Milky Way disk and are suffering from high extinctions, thus a uniform prior
  is not suitable.

  The temperatures and radii obtained by the SED fitting are basically
  consistent with the \Gaia DR3 results. The SED fitting of temperature and
  radius for all targets are in agreement with \Gaia's results within a $1.7 \sigma$
  confidence interval. The median difference for temperature is 115 K and for
  radius is 0.1 $\Rsun$. J0709 has the largest measurement difference with \Gaia's
  result, with a temperature difference of 438 K and a radius difference of 0.43 $\Rsun$.
  In the SED fitting of all the selected objects, the models are well-matched
  with the observed SEDs, and no obvious IR excess is found.

  \begin{figure*}
    \includegraphics[width=0.99\textwidth]{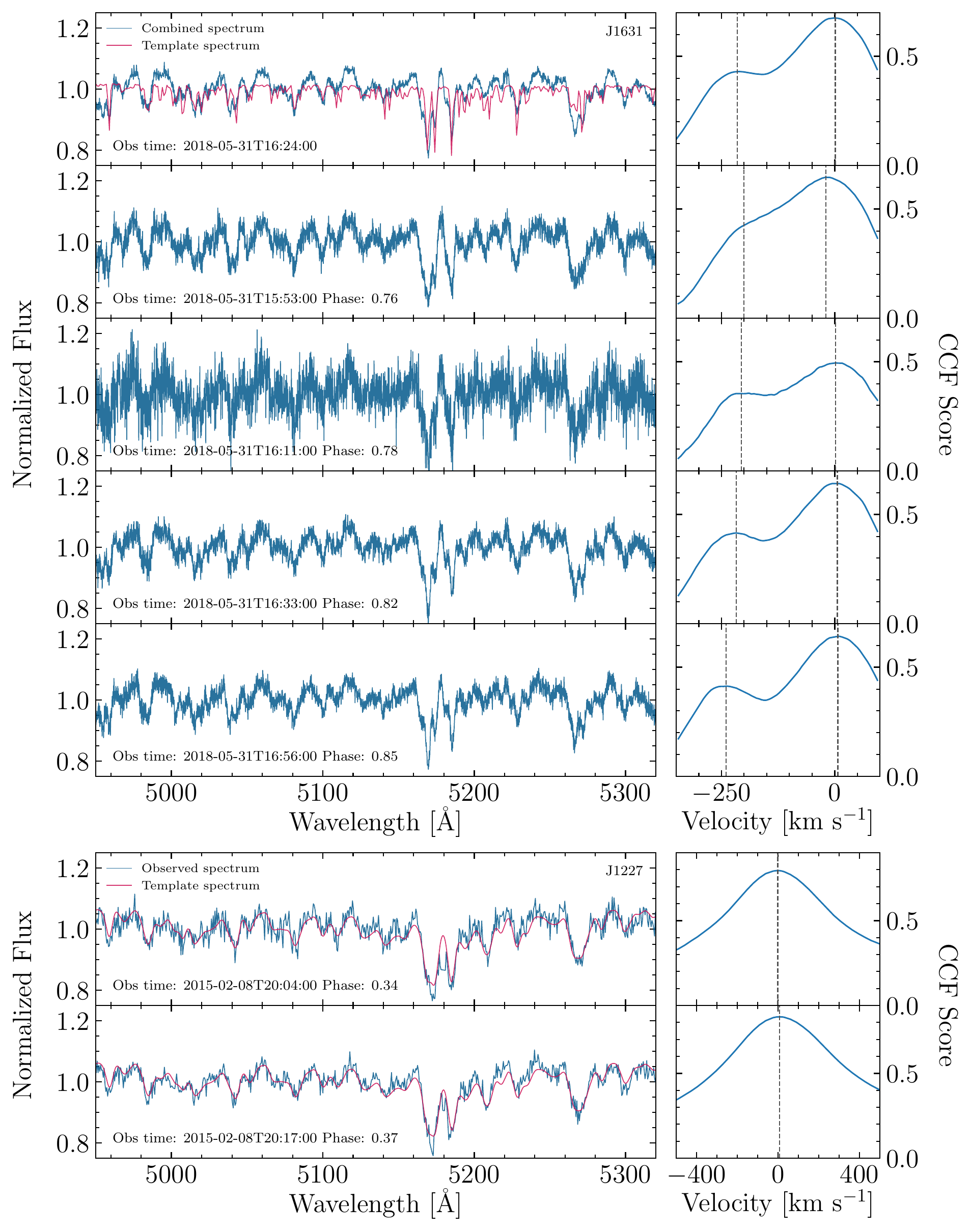}
    \caption{LAMOST spectra (left panels) and CCF (right panels) for J1631 (upper panels)
      and J1227 (lower panels). The uppermost panel in J1631 shows the combined spectrum (blue)
      and a template spectrum (red) used to measure the CCF profile.
      The dashed lines mark the peaks in the CCF panels.
      It is evident from the CCF that J1631 is likely to be a double-lined spectroscopic binary, while J1227 is a single-lined spectroscopic binary.
      The UTC time for each observation and the corresponding orbital phase is indicated in the figure legend.
      \label{fig:LAMOST_Spectrum_for_Two_targets}}
  \end{figure*}

  \subsection{LAMOST spectroscopy}

  The LAMOST survey provides both medium-resolution spectra (MRS) and
  low-resolution spectra (LRS). Typically, two to four consecutive exposures
  (10-20 minutes per exposure) are taken for each observation.
  Both single and combined spectra are available. We found two of ten
  objects have LAMOST observations.
  One of the objects, J1631\footnote{\Gaia Source ID: 1410477980245374848.
    Thereafter, we will use short names to distinguish each target.
    The \Gaia source ID of each target and the corresponding short name are listed in Table \ref{table:1}.},
  was observed by LAMOST MRS on 2018 May 31 with four consecutive exposures.
  The other object, J1227, was observed by LAMOST LRS on 2015 Feb 08 with two exposures.

  We use the cross-correlation function (CCF) to measure the radial velocities
  of the LAMOST spectra. The templates are interpolated using \texttt{The Payne}\footnote{\url{https://github.com/pacargile/ThePayne}},
  a spectral interpolation tool based on a neural-net algorithm.
  We use the BOSZ grid \citep{2017AJ....153..234B} to train the interpolated model.
  The effective temperatures provided by \Gaia DR3 are used as prior
  to choose the model template. The template spectra have been broadened
  to match the spectral resolution (R=7500) of the observed
  spectra before the training. We find the best stellar parameters
  by minimizing the $\chi^2$ of the templates and the observed spectra.

  The best fitting stellar parameters of J1631 is
  $\Teff = 5700$\,K, ${\rm [Fe/H]} = -1.5$. The corresponding template is
  used to measure the RVs of the LAMOST spectra. We find that the
  template could not match the observed spectra well, and the CCF
  structures show double peak structures, which indicates that the
  system could be a SB2 (Figure \ref{fig:LAMOST_Spectrum_for_Two_targets}). For the
  four LAMOST MRS spectra (each exposure time is 20 minutes), we find
  that the strongest peak on the CCFs fell to near zero and did not move
  significantly. The second peak has an RV of about $-250$ \kms.

  To evaluate whether J1631 is a truly SB2, we decompose the spectra
  of J1631 to obtain the temperatures and RVs of the system more
  precisely. We use a method similar to \citet{elbadry2018} to
  construct a spectral model for binary systems. The model contains
  two radiative components, and the two spectra can be adjusted
  by their corresponding RVs and the flux ratio of the two components.
  We use Markov chain Monte Carlo (MCMC) method to fit the parameters of the binary spectral
  model. We assume that the stars of the binary have the same origin
  and elemental abundance. Therefore, only one set of elemental
  components needs to be retained in the model. For simplicity,
  we do not introduce the stellar evolution model in the fitting
  to constrain the stellar parameters. Finally, our model includes
  ten free parameters, including $T_\mathrm{eff,1}$, $T_\mathrm{eff,2}$,
  $\log g_1$, $\log g_2$, $\rm [M/H]$, $\rm [C/M]$, $\rm [\alpha/M]$,
  $V_{\rm r,1}$, $V_{\rm r,2}$, and the flux ratio of the two spectral
  components.

  Figure \ref{fig:Spectrum_Decomposition} shows the best fitting result of the combined
  spectrum, which gives the temperatures of the two components of
  $5931 \pm 23$ K and $5870 \pm 30$ K. The metal abundance is ${\rm [M/H]} = -0.26$.
  We fixed the stellar parameters to the best fitting results of the
  combined spectrum and then measured the RVs of each single spectrum.
  The RVs for the higher temperature component measured in four consecutive
  spectra are $-$202 \kms, $-$212 \kms, $-$213 \kms, and $-$216 \kms.
  Another lower temperature component has RVs of $-$2 \kms,
  6 \kms, 6 \kms, and 8 \kms.

  For J1227, only the LAMOST LRS survey observed this object twice with
  an exposure time of 15 minutes each. The template matches the observed
  spectra well, and the RVs of the two spectra are $7$ \kms
  and $-1$ \kms, respectively.

  \section{Analyzes} \label{sec:discuss}

  The ten targets we selected in Section \ref{sec:data} can be classified
  into three categories according to their respective observations:
  two targets (J1227 and J1631, labelled with L in Table \ref{table:1})
  have LAMOST spectra; seven targets (J0056, J0216, J0952, J1227,
  J1631, J2023, and J2305, labelled with P) which have folded light curve periods from \TESS
  shorter than the theoretical minimum orbital period; and the
  remaining three targets (J0623, J0709, and J0805, labelled with C for the first two targets)
  have no apparent inconsistencies in the parameters. We discuss the three types of
  targets in the following.

  \subsection{Spectroscopy} \label{sec:spectro}

  High-quality spectra can be used to estimate whether there are multiple
  luminous components. To validate the SB1 nature and exclude
  false positives (potential SB2) of our sample, we check the CCF profiles
  of each available LAMOST spectrum, by correlating with a best-fitted template.
  If two or multiple components with similar flux contributions are presented
  in a high-resolution spectrum, the CCF will feature double or even multiple
  peaks. Otherwise, if the system has only one dominant component, the CCF will
  show a single peak, indicating a good correlation between the template and the observation.

\begin{figure*}
    \includegraphics[width=0.99\textwidth]{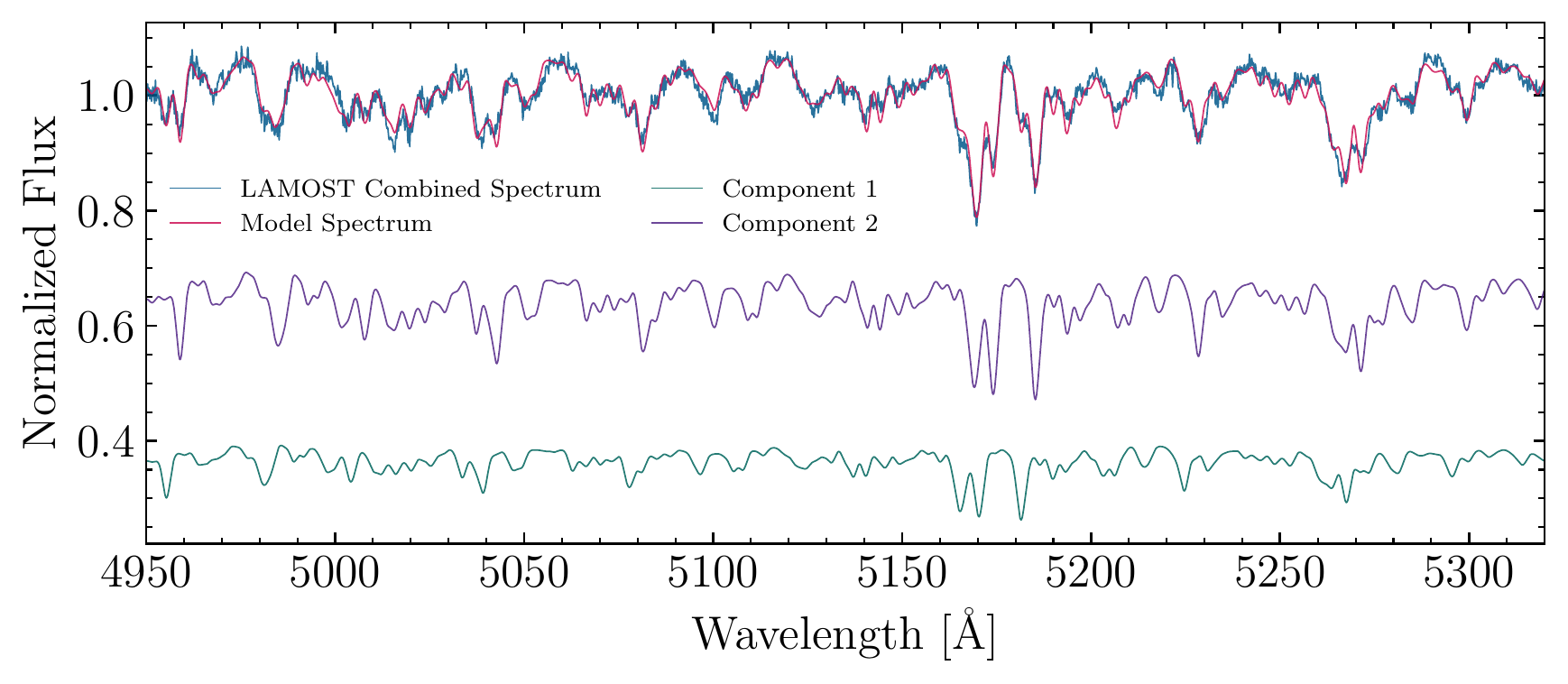}
    \caption{Decomposition of the combined spectrum of J1631.
      \label{fig:Spectrum_Decomposition}}
  \end{figure*}

  We show the LAMOST spectra of the two targets
  and the corresponding template spectrum used to measure the CCF profile in Figure \ref{fig:LAMOST_Spectrum_for_Two_targets}.
  For J1631, we found that the template cannot match the observed spectra well
  and the CCFs show clear double peaks. We decomposed the combined
  spectrum of J1631. The good spectral decomposition (see Figure \ref{fig:Spectrum_Decomposition})
  demonstrates our statement that J1631 is a SB2. As the object has only been observed
  for less than one hour, we are unable to determine the mass ratio of the binary.

  J1227 has two LAMOST LRS spectra, and the template matches well with the observed spectra.
  Given the current observations, the single-peak feature of both CCFs suggests that J1227 is likely to be a SB1.
  However, the measured RVs for the two spectra are close to zero, the reason for which will be discussed more in Section~\ref{sec:phase}.
  We stress that more spectroscopic observations with higher spectral resolution and phase coverage are needed
  to validate the nature of J1227.

  \begin{figure*}
    \includegraphics[width=0.99\textwidth]{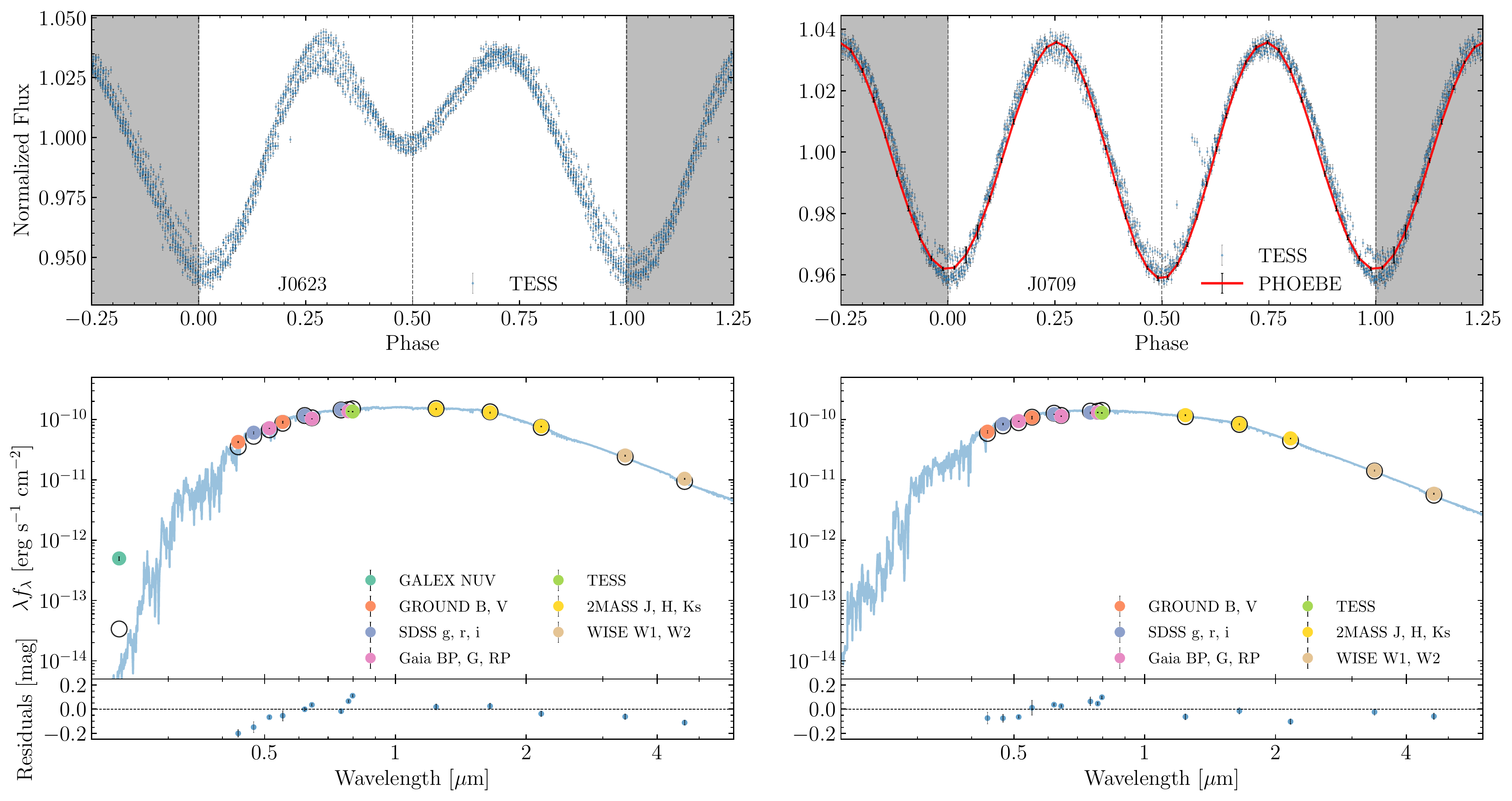}
    \caption{The light curves and SED fittings of the two targets with
      candidate compact objects (J0623 and J0709). Note that the light
      curves beyond phase 0 - 1 are repeated to better show their ellipsoidal
      shape, and aligned to ensure the primary minima are in phase 0.
      The observed \TESS light curves and the modeled light curves by
      PHOEBE are shown in blue dots and red lines, respectively. The observed
      values from the same survey are marked by the same color in the
      SED fitting panels, and the modeled values are marked with black circles.}
    \label{fig:SED_Fitting}
  \end{figure*}

  \subsection{Photometric period and orbital phase} \label{sec:phase}

  Seven targets in Table \ref{table:1} have photometric periods shorter
  than their theoretical minimum period. This indicates a non-physical
  picture that the radii of the visible stars in these binary systems
  exceed their Roche radii. One possible reason is that these systems
  actually consist of two main sequence stars, which are treated as a
  single star in the pipeline of \Gaia to measure the stellar parameters.
  The result is an overestimated stellar radius and hence an overestimated
  theoretical minimum orbital period. In addition, the photometric period
  obtained from \TESS is not necessarily the system's orbital period.
  It is possible that the photometric variation could have originated
  from intrinsic variability such as stellar rotation. Due to the
  lack of additional observations, we cannot rule out these possibilities.

  Two targets have spectral observations from LAMOST. Given their orbital ephemeris,
  we calculated and showed the orbital phase of each LAMOST spectra in Figure \ref{fig:LAMOST_Spectrum_for_Two_targets}.
  The zero points are set to the same as the primary minima of the light curve. Therefore,
  phase 0.5 corresponds with the scenario that the radial velocities of the two components
  are both zero.

  The LAMOST spectra of J1631 were taken in May 2018, with a two-year gap to \TESS observations.
  In order to calculate the orbital phases at which the spectra were taken,
  we use \TESS Sector 24 and Sector 25 (from April to June in 2020) in conjunction to achieve
  a longer observation time span (than just one sector) and therefore to obtain an orbital
  period with better accuracy. According to our calculation, the four spectra were taken
  at times corresponding to phases 0.76 - 0.85, which is the case when the radial velocity
  difference between the two stars is near the maximum. The result is consistent with the
  double peak shown in the CCF profile.

  The radial velocity of J1227 cannot be aligned with its orbital ephemeris.
  Given the radial velocity variation $\DRV \simeq 727$ \kms measured by \Gaia, it is expected that J1227
  should embody a radial velocity variation $\DRV \simeq 42$ \kms between the two LAMOST
  observations, which is correspondingly at phase 0.34 and 0.37. However, CCF measured radial
  velocity variation $\DRV \simeq 8$ \kms disagrees with the expected variation.
  Considering that only one \TESS Sector (Sector 22 in 2020) has monitored J1227,
  which leaves a large time gap of five years with respect to LAMOST observations (in February 2015),
  the calculated phase at which the LAMOST spectra were taken could therefore suffer
  from much larger uncertainty due to possible inaccurate period.
  Photometric observations with a longer time span will help to reduce
  the uncertainty of period folding and better determine the phases of spectra.
  Moreover, J1227 could consist of two stars with similar temperatures
  and luminosities. Since J1227 has only low-resolution spectra
  taken by LAMOST, the superposition of the two components' spectra
  with opposite RVs will lead to RV smearing on the observed spectra, resulting
  in CCF measurements yielding two near-zero RVs.

  \subsection{Promising targets}

  There are two bright stars next to J0805. Due to the interference of
  these bright stars, we cannot obtain reliable photometric magnitudes
  of this object, and thus cannot fit the SED. Because of the concern
  that the photometric magnitudes and the light curves may be contaminated
  by the surrounding bright stars, we excluded this object from the candidates.

  The two remaining targets (J0623 and J0709) in Table \ref{table:1} satisfy all the criteria
  listed in Section \ref{sec:data}. They have large radial velocity variations,
  ellipsoidal-like light curves, and their temperatures and radii given by
  SED fitting are in agreement with the \Gaia measurements within a $1.7\sigma$ confidence interval (Figure \ref{fig:SED_Fitting}).
  The lack of X-ray observations queried from HEASARC\footnote{\url{https://heasarc.gsfc.nasa.gov/xamin/xamin.jsp}}
  indicates that J0709 is quiescent. J0623 has X-ray detections by XMM-Newton \citep{2008A&A...480..611S}
  with a flux range of $2.714 \times 10^{-12} \sim 3.817 \times 10^{-12} \ {\rm erg} \ {\rm s}^{-1} \ {\rm cm}^{-2}$.

\begin{figure*}
    \centering
    \includegraphics[width=0.85\textwidth]{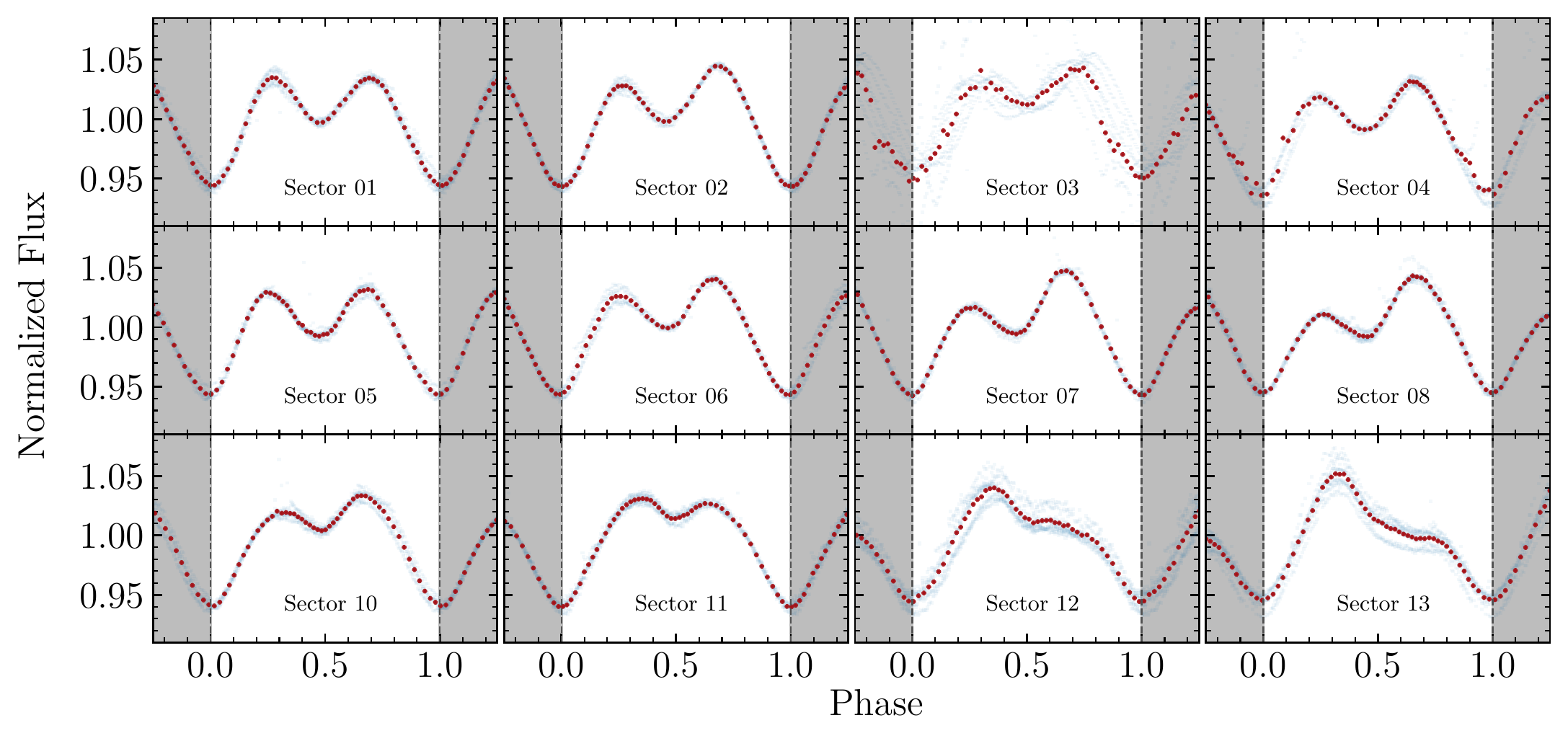}
    \caption{The \TESS light curves of J0623 in multiple sectors. The shallow
      blue dots show the original normalized flux, and the red dots represent the
      mean flux of each bin.
      \label{fig:Multiple_Sectors_for_Lightcurve_of_J0623}}
  \end{figure*}

  We use PHOEBE to fit the light curves of possible pure ellipsoidal targets.
  As shown in Figure \ref{fig:SED_Fitting}, J0709 can be well-fitted by the pure ellipsoidal light curve.
  However, the light curve of J0623 shows a noticeable difference in depth between its two minima.
  The two peaks of the light curve are also slightly different, which the pure ellipsoidal model
  can hardly explain. The observations in multiple TESS sectors show a highly asymmetry and variable
  light curve (Figure \ref{fig:Multiple_Sectors_for_Lightcurve_of_J0623}). This may indicate the existence of stellar activity on
  the surface of J0623, and the light curve shape of J0623 could be due to a combination of
  ellipsoidal variability and stellar activity. The light curve of J0623 is similar to a recently
  reported white dwarf-main-sequence binary system \citep{2022ApJ...936...33Z}. These two targets contain
  potential compact objects, and deserve further constraints on their stellar and orbital parameters
  through follow-up observations.

  \begin{figure*}
    \centering
    \includegraphics[width=0.85\textwidth]{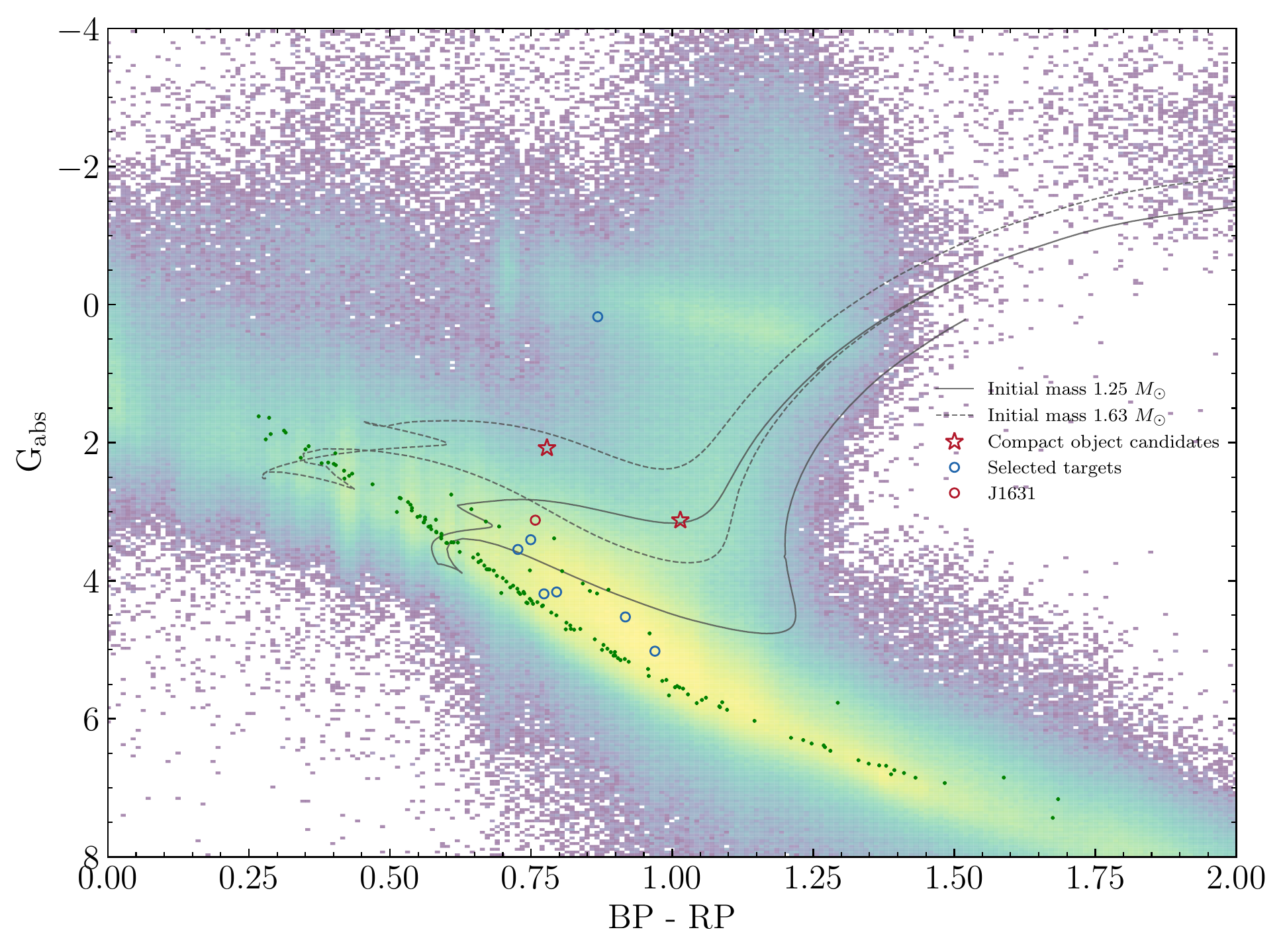}
    \caption{The color-magnitude diagram of the ten targets in Table \ref{table:1}.
      The extinction of G-band magnitude and the BP-RP color reddening have been corrected.
      The solid and dashed lines show the single star evolution tracks with respective
      initial masses of 1.25 and 1.63 Solar mass, as suggested by \Gaia's measurement.
      For reference, we plot members of Hyades and Praesepe clusters, using green dots.
      These members form a clear single-star sequence and a relatively sparse binary-star
      sequence which lies $\sim$0.75 magnitude above.
      \label{fig:Color_Magnitude_Diagram}}
  \end{figure*}

  \subsection{Radial velocity}

  Our selection criteria require $\massfunc > 1 \Msun$, which prioritized
  those targets with large radial velocity variations ($\DRV > 200$ km s$^{-1}$).
  \citet{2022arXiv220605486B} show that the radial velocity
  measurements of \Gaia have an average error of $|\sigma_{\Vr}| < 3$ km s$^{-1}$
  for targets with $\Teff < 6000$ K. \citet{2022arXiv220605902K}
  present additional bounds for determining whether a target
  is truly velocity variable. Our criteria fully satisfy these requirements,
  and it is reasonable to assume that the velocity variation of
  our samples is credible. However, LAMOST spectra suggest us excluding
  J1631, which clearly shows double peaks in its CCF profile.
  In order to exclude false-positive samples like SB2, and to constrain more
  precisely on orbital parameters in conjunction with light curves, either
  the original \Gaia spectrum or multiple follow-up spectroscopic observations
  are required.

  \subsection{UV excess}

  Four of the ten objects (J0056, J0216, J0623, and J1631) have GALEX \citep{2005ApJ...619L...1M} ultraviolet (UV) observations,
  all of which show near-UV (NUV) excess, that is, the observed NUV flux is higher than the theoretical SED model.
  For instance, J0623 has an NUV flux of about one magnitude of order larger than the model
  (e.g., see the left bottom panel of Figure \ref{fig:SED_Fitting}).
  We offer the following two possible indications.
  (1) Chromospheric activities \citep{2017ARA&A..55..159L} of the visible star might contribute to a substantial amount of UV excess emission,
  which the model SED does not contain (model SED contains only the photospheric emissions).
  The presence of the chromospheric UV emission is supported by the fact that the visible star of J0623 is a spotted rotational star (Figure \ref{fig:Multiple_Sectors_for_Lightcurve_of_J0623}),
  evidence of wild stellar activities.
  (2) An accretion process due to the mass transfer from the donor to the compact object might emit UV radiation.
  This scenario often happens in close binary systems.
  We stress that however, the conjecture of such an accretion might need to be further examined,
  using follow-up spectroscopy (i.e., to check whether there is evidence of H$\alpha$ emission line of He emission line by an accretion disk).

  \subsection{Color-magnitude diagram}

  We show the color-magnitude diagram of the ten targets in Table \ref{table:1} and 0.28 million
  randomly selected targets from \Gaia as background in Figure \ref{fig:Color_Magnitude_Diagram}. The color index and
  extinction-corrected G-band magnitude are provided by \Gaia. The topmost target J0805 is situated
  in the giant branch, which may be due to the contamination from the nearby bright stars.

  There are seven targets located in the main sequence. We retrieve the color index
  and magnitude of members of Hyades and Praesepe clusters to distinguish the systems in single
  star main sequence and others in binary star main sequence \citep{2018A&A...616A..10G}.
  Four targets (J0216, J1631, J2023, and J2305) are closer with the binary star main sequence.
  Noteworthy, J1631, a spectroscopically confirmed SB2, is located nearby the binary star main sequence.
  These possible binary star systems may support one of our explanations about the orbital period
  conflict: some binary star systems were treated as single star systems in \Gaia's pipeline, thus
  leading to larger radius measurements.

  The two targets contain compact object candidates are located on the fringe of the main sequence.
  Considering they have magnitude exceeds of about 1 - 2 dex above the main sequence, their deviations
  from the main sequence can hardly be explained by binary star systems. Moreover, the two targets
  are closer to the subgiant phase in \texttt{MIST} evolution tracks, where the initial masses
  are set to the same as the \Gaia measurements. The two targets are more likely to be subgiants
  which have recently evolved and detached from the main sequence.

  \section{Conclusions and discussion} \label{sec:conclusion}

  In this paper, we try to search for potential compact binary systems
  from the newly released \Gaia DR3 database. Based on the radial
  velocity variations $\DRV$, the masses $M_1$ and radii $R_1$ of the visible stars
  provided by \Gaia, we have selected a sample of 65 sources with mass
  functions $\massfunc > 1 \Msun$. Ten of these sources have ellipsoidal-like
  light curves in \TESS. The properties of these samples
  are summarized in Table \ref{table:1}.

  Utilizing LAMOST spectroscopy available for two of our targets,
  we exclude one false positive J1631 as a SB2. The other target J1227
  is likely to be a SB1 but shows no significant RV variation between two
  consecutive exposures, therefore the nature of which remains uncertain.
  J1227 could potentially be a false positive. Although we selected the
  sources according to criteria very similar to \citet{2022arXiv220605902K},
  we still could not completely exclude the potential false-positive
  targets in the selected samples. Therefore, for targets with large radial
  velocity variations in \Gaia, we stress that it is necessary to perform a
  double check by utilizing spectroscopy from other surveys like LAMOST,
  or from follow-up observations, to draw a more convincing conclusion.

  There are seven targets in Table \ref{table:1}, for which the theoretical
  minimum orbital periods is larger than the periods calculated
  by folding the corresponding \TESS light curves. This contradiction
  may be attributed to the fact that these sources are in fact binary
  star systems, while the mass and radius in \Gaia are mistakenly
  processed as a single star. We also cannot fully confirm that the
  folding period of the \TESS light curve is the true orbital period
  of the binary system.

  Finally, two targets (J0623 and J0709) satisfy all
  the criteria described in Section \ref{sec:data}, and their parameters are more
  consistent with single-lined spectroscopic binaries. We consider them as potential compact binary systems.
  These two are potential compact object candidates and are worth follow-up
  observations to constrain their stellar parameters. As an illustration,
  we only study samples with $\massfunc > 1 \Msun$. More worthy targets may be
  found after loosening the requirement of the mass function.
  Compared with similar works which dedicated to unveiling compact objects from \Gaia
  classified SB1 targets \citep[e.g.,][]{elbadry2022,2022arXiv220705086J}, we investigate
  the \Gaia DR3 main table which contains $\sim$1.8 billion objects, thus enabling
  us to explore more potential candidates. Moreover, utilizing other spectroscopic observations
  such as LAMOST can prevent our samples from contaminated by false positives.
  Our work can serve as an example of the dynamical method to search for
  compact objects in quiescent and demonstrates the potential to
  filter more compact objects from \Gaia.

  \begin{acknowledgments}
    We thank Mouyuan Sun, Qian-Yu An, Hao-Yan Chen, Rui Wang, Jia-Yi Chen,
    Hao-Bin Liu, and Song Wang for helpful discussions. We thank the anonymous referee
    for constructive suggestions that improved the paper.
    This work was supported by the National Key R\&D Program of China under
    grant 2021YFA1600401, the National Natural Science Foundation of China under grants
    11925301, 11933004, 11988101, 12033006, 12103041, and 12221003, and the China Postdoctoral Science
    Foundation under grant 2021M702742. Guoshoujing Telescope (the Large
    Sky Area Multi-Object Fiber Spectroscopic Telescope, LAMOST) is a National Major Scientific
    Project built by the Chinese Academy of Sciences. Funding for the project has been provided
    by the National Development and Reform Commission. LAMOST is operated and managed by the
    National Astronomical Observatories, Chinese Academy of Sciences.
  \end{acknowledgments}

  \software{
    astroARIADNE \citep{2022MNRAS.tmp..920V},
    Astropy \citep{2013A&A...558A..33A,2018AJ....156..123A},
    Lightkurve \citep{2018ascl.soft12013L},
    Matplotlib \citep{Hunter:2007},
    MIST \citep{2016ApJS..222....8D},
    NumPy \citep{harris2020array},
    pandas \citep{mckinney-proc-scipy-2010},
    PHOEBE \citep{2020ApJS..250...34C},
    The Payne \citep{ting2019}
  }

  \begin{longrotatetable}
    \begin{deluxetable*}{clccccccccccccccc}
      \tablecaption{Characteristics of selected targets from \Gaia DR3 \label{table:1}}
      \tablewidth{0.99\textwidth}
      \tabletypesize{\scriptsize}
      \tablehead{
        \colhead{ID} & \colhead{\Gaia Source ID} & \colhead{R.A.} & \colhead{Dec} & \colhead{$M_1$}     & \colhead{$R_1$}     & \colhead{$\DRV$}        & \colhead{$\Teff$} & \colhead{$P_{\rm orb}^{\rm min}$} & \colhead{$P_{\rm orb}^{\rm lc}$} & \colhead{$\massfunc_{\rm Gaia}$} & \colhead{$\massfunc_{\rm lc}$} & \colhead{$T_{\rm eff,SED}$}   & \colhead{$R_{\rm 1,SED}$} & \colhead{$BP-RP$} & \colhead{$G_{\rm abs}$} & \colhead{Label} \\
        \colhead{}   & \colhead{}                & \colhead{}     & \colhead{}    & \colhead{(1)}       & \colhead{(2)}       & \colhead{(3)}           & \colhead{(4)}     & \colhead{(5)}                     & \colhead{(6)}                    & \colhead{(7)}                    & \colhead{(8)}                  & \colhead{(9)}                 & \colhead{(10)}            & \colhead{(11)}    & \colhead{(12)}          & \colhead{(13)} \\
        \colhead{}   & \colhead{}                & \colhead{}     & \colhead{}    & \colhead{($\Msun$)} & \colhead{($\Rsun$)} & \colhead{(km s$^{-1}$)} & \colhead{(K)}     & \colhead{(days)}                  & \colhead{(days)}                 & \colhead{($\Msun$)}              & \colhead{($\Msun$)}            & \colhead{(K)}                 & \colhead{($\Rsun$)}       & \colhead{}        & \colhead{}              & \colhead{}
      }
      \startdata
      J0056 & 4987200236093193216 & 00:56:18.0 & -41:21:56.45 & 0.963 & $1.30_{-0.08}^{+0.03}$ & 603.115 & $5704_{-19}^{+70}$ & 0.559 & 0.334 & 1.59 & 0.95 & $5819_{-262}^{+203}$   & $1.26_{-0.11}^{+0.11}$ & 0.795 & 4.16 & P \\
      J0216 & 4944640992840656000 & 02:16:25.3 & -43:12:10.58 & 1.106 & $1.58_{-0.01}^{+0.02}$ & 616.590 & $5927_{-27}^{+35}$ & 0.697 & 0.416 & 2.12 & 1.26 & $5919_{-256}^{+247}$   & $1.62_{-0.10}^{+0.12}$ & 0.727 & 3.55 & P \\
      J0623 & 5284413278650095872 & 06:23:34.3 & -66:17:18.99 & 1.251 & $2.74_{-0.14}^{+0.08}$ & 452.383 & $5061_{-26}^{+47}$ & 1.497 & 3.562 & 1.79 & 4.27 & $4798_{-131}^{+135}$   & $3.08_{-0.20}^{+0.20}$ & 1.015 & 3.12 & C \\
      J0709 & 5602540372393859968 & 07:09:01.9 & -34:51:56.10 & 1.627 & $3.31_{-0.01}^{+0.01}$ & 393.256 & $5918_{-6}^{+6}$   & 1.739 & 3.513 & 1.37 & 2.77 & $5480_{-248}^{+248}$   & $3.74_{-0.28}^{+0.39}$ & 0.778 & 2.08 & C \\
      J0805 & 5546086050952797952 & 08:05:09.5 & -33:33:04.76 & 2.401 & $9.20_{-0.16}^{+0.25}$ & 242.124 & $5575_{-11}^{+9}$  & 6.638 & 7.019 & 1.22 & 1.29 & \nodata                & \nodata                & 0.868 & 0.18 & \nodata \\
      J0952 & 5687390848640176000 & 09:52:58.8 & -13:48:29.50 & 0.991 & $1.23_{-0.01}^{+0.01}$ & 590.239 & $5798_{-4}^{+3}$   & 0.504 & 0.450 & 1.34 & 1.20 & $5449_{-382}^{+224}$   & $1.38_{-0.10}^{+0.17}$ & 0.773 & 4.19 & P \\
      J1227 & 1542461401838152960 & 12:27:29.4 & +47:44:19.35 & 0.906 & $1.12_{-0.03}^{+0.01}$ & 726.781 & $5128_{-12}^{+11}$ & 0.461 & 0.273 & 2.29 & 1.36 & $4876_{-333}^{+175}$   & $1.22_{-0.15}^{+0.11}$ & 0.970 & 5.02 & LP \\
      J1631 & 1410477980245374848 & 16:31:55.3 & +47:35:10.51 & 1.193 & $1.88_{-0.02}^{+0.02}$ & 472.610 & $5936_{-7}^{+7}$   & 0.872 & 0.476 & 1.19 & 0.65 & $5836_{-132}^{+181}$   & $1.98_{-0.10}^{+0.11}$ & 0.758 & 3.12 & LP \\
      J2023 & 2243723836359839872 & 20:23:48.9 & +63:33:43.03 & 0.908 & $1.27_{-0.01}^{+0.02}$ & 579.668 & $5340_{-78}^{+51}$ & 0.555 & 0.301 & 1.40 & 0.76 & $5288_{-210}^{+192}$   & $1.29_{-0.09}^{+0.09}$ & 0.917 & 4.53 & P \\
      J2305 & 1931538743969686912 & 23:05:51.3 & +42:32:43.32 & 1.116 & $1.76_{-0.04}^{+0.04}$ & 611.487 & $5825_{-39}^{+42}$ & 0.812 & 0.428 & 2.40 & 1.27 & $5789_{-172}^{+265}$   & $1.79_{-0.19}^{+0.10}$ & 0.748 & 3.41 & P \\
      \enddata
      \tablecomments{
        (1 - 4) The mass ($M_1$), radius ($R_1$), radial velocity variation ($\DRV$),
        and temperature ($\Teff$) measured by \Gaia. (5) The theoretical minimum orbital
        period ($P^{\rm min}_{\rm orb}$) calculated from the \Gaia parameters. (6) The photometric
        period ($P^{\rm lc}_{\rm orb}$) folded from the \TESS light curve. (7) The mass function
        of the invisible components ($\massfunc_{\rm Gaia}$) calculated from $P^{\rm min}_{\rm orb}$. (8) The
        mass function of the invisible components ($\massfunc_{\rm lc}$) calculated from $P^{\rm lc}_{\rm orb}$.
        (9 - 10) The temperature ($T_{\rm eff,SED}$) and radius ($R_{\rm 1,SED}$) measured by SED fitting. (11 - 12) The color
        index ($B_{\rm p} - R_{\rm p}$) and G band magnitude ($G_{\rm abs}$) measured by \Gaia.
        (13) The label used to separate targets into different categories:
        C for a probably compact object candidate, L for a target with LAMOST
        spectroscopic observations, and P for a target with a photometric period
        shorter than the theoretical minimum orbital period.}
    \end{deluxetable*}
  \end{longrotatetable}

  \appendix
  \section{Full Selected Samples}

  \startlongtable
  \begin{deluxetable*}{ccccccccccc}
    \tablecaption{Characteristics of 65 selected targets from \Gaia DR3 \label{suptable}}
    \tablewidth{0.99\textwidth}
    \tabletypesize{\scriptsize}
    \tablehead{
      \colhead{\Gaia Source ID} & \colhead{R.A.} & \colhead{Dec} & \colhead{$M_1$}     & \colhead{$R_1$}     & \colhead{$\DRV$}        & \colhead{$\Teff$} & \colhead{$P^{\rm min}_{\rm orb}$} & \colhead{$P^{\rm lc}_{\rm orb}$} & \colhead{$\massfunc_{\rm Gaia}$} & \colhead{$\massfunc_{\rm lc}$} \\
      \colhead{}                & \colhead{}     & \colhead{}    & \colhead{(1)}       & \colhead{(2)}       & \colhead{(3)}           & \colhead{(4)}     & \colhead{(5)}                     & \colhead{(6)}                    & \colhead{(7)}                    & \colhead{(8)}         \\
      \colhead{}                & \colhead{}     & \colhead{}    & \colhead{($\Msun$)} & \colhead{($\Rsun$)} & \colhead{(km s$^{-1}$)} & \colhead{(K)}     & \colhead{(days)}                  & \colhead{(days)}                 & \colhead{($\Msun$)}              & \colhead{($\Msun$)}
    }
    \startdata
    431341435235181056  & 00:13:09.1 & +63:09:20.27 & 4.687 &  33.892 &  169.310 & 5879 &  33.616 &  40.000 &   2.11 &    2.51 \\
    524536457422822016  & 00:51:23.5 & +65:13:05.60 & 5.333 &   4.479 &  487.597 & 5762 &   1.514 &   2.326 &   2.27 &    3.49 \\
    423647194605271808  & 00:52:08.6 & +55:35:43.56 & 1.142 &   8.148 &  223.900 & 4417 &   8.028 &   5.955 &   1.17 &    0.87 \\
    4987200236093193216 & 00:56:18.0 & -41:21:56.45 & 0.963 &   1.302 &  603.115 & 5704 &   0.559 &   0.334 &   1.59 &    0.95 \\
    297802831758073984  & 01:58:29.6 & +26:03:33.13 & 1.140 &   1.922 &  496.225 & 5675 &   0.921 &   0.455 &   1.46 &    0.72 \\
    346272671566816512  & 02:03:09.3 & +42:36:51.00 & 2.501 &   7.509 &  304.104 & 5209 &   4.799 &  20.756 &   1.75 &    7.56 \\
    508371746710670976  & 02:04:55.0 & +60:55:45.07 & 6.112 &   4.694 &  654.221 & 5982 &   1.517 &  28.465 &   5.50 &  103.22 \\
    4944640992840656000 & 02:16:25.3 & -43:12:10.58 & 1.106 &   1.581 &  616.590 & 5927 &   0.697 &   0.416 &   2.12 &    1.26 \\
    516035548914422400  & 02:30:10.9 & +64:58:32.51 & 1.085 &   1.689 &  609.841 & 5723 &   0.777 &   0.391 &   2.28 &    1.15 \\
    5079746164163047296 & 03:03:13.4 & -20:36:50.98 & 1.056 &   1.228 &  610.946 & 5917 &   0.489 &   0.335 &   1.44 &    0.99 \\
    5099527782799972736 & 03:16:12.0 & -22:05:30.87 & 1.097 &   1.868 &  565.268 & 5589 &   0.899 &   0.772 &   2.10 &    1.81 \\
    4853820950132560896 & 03:17:55.4 & -39:16:48.76 & 1.095 &   1.501 &  530.947 & 5955 &   0.648 &   1.650 &   1.26 &    3.20 \\
    5086274102138599808 & 03:33:07.0 & -24:17:00.10 & 0.965 &   1.391 &  542.507 & 5636 &   0.616 & \nodata &   1.27 & \nodata \\
    2915702118705099520 & 05:53:02.8 & -23:22:28.60 & 1.057 &   1.260 &  582.941 & 5891 &   0.507 &   0.337 &   1.30 &    0.86 \\
    3344011302729909760 & 06:09:17.0 & +12:32:48.48 & 3.396 &  15.443 &  205.060 & 5169 &  12.147 &  30.046 &   1.36 &    3.36 \\
    5284413278650095872 & 06:23:34.3 & -66:17:18.99 & 1.251 &   2.741 &  452.383 & 5061 &   1.497 &   3.562 &   1.79 &    4.27 \\
    2922949687038297088 & 06:33:30.0 & -26:22:16.05 & 0.993 &   1.333 & 1083.699 & 5775 &   0.570 &   0.679 &   9.39 &   11.19 \\
    2928949511533366784 & 07:03:14.2 & -21:21:47.26 & 0.929 &   1.305 &  608.021 & 5530 &   0.571 &   0.411 &   1.66 &    1.20 \\
    5602540372393859968 & 07:09:01.9 & -34:51:56.10 & 1.627 &   3.307 &  393.256 & 5918 &   1.739 &   3.513 &   1.37 &    2.77 \\
    5275788125325011584 & 07:47:57.6 & -64:44:23.35 & 1.080 &   2.979 &  710.641 & 4925 &   1.825 &   2.227 &   8.48 &   10.35 \\
    5546086050952797952 & 08:05:09.5 & -33:33:04.76 & 2.401 &   9.195 &  242.124 & 5575 &   6.638 &   7.019 &   1.22 &    1.29 \\
    707471754641398144  & 08:22:35.5 & +28:55:18.70 & 1.070 &   4.244 &  317.901 & 4804 &   3.118 & \nodata &   1.30 & \nodata \\
    1123617153201142656 & 08:25:58.6 & +74:51:01.39 & 1.103 &   1.311 &  572.739 & 5993 &   0.527 &   0.417 &   1.28 &    1.02 \\
    5709247009497151872 & 08:38:14.9 & -16:58:31.49 & 1.050 &   1.727 &  709.356 & 5603 &   0.817 &   0.662 &   3.78 &    3.06 \\
    5709342980542992768 & 08:45:05.3 & -16:25:02.77 & 0.900 &   1.040 &  729.708 & 5577 &   0.413 &   0.665 &   2.08 &    3.34 \\
    5325906785898343168 & 09:10:21.0 & -49:23:24.86 & 1.597 &   3.247 &  415.958 & 5572 &   1.708 &   0.767 &   1.59 &    0.71 \\
    714884906850837120  & 09:16:12.2 & +36:15:33.34 & 0.842 &   1.211 &  585.521 & 4719 &   0.536 &   0.367 &   1.39 &    0.95 \\
    5435132408441359360 & 09:47:54.0 & -35:17:26.66 & 0.931 &   1.356 &  664.705 & 5516 &   0.604 &   0.327 &   2.30 &    1.24 \\
    5687390848640176000 & 09:52:58.8 & -13:48:29.50 & 0.991 &   1.228 &  590.239 & 5798 &   0.504 &   0.450 &   1.34 &    1.20 \\
    5200508962219751424 & 11:26:40.6 & -77:10:14.43 & 6.617 & 178.770 &  501.350 & 3315 & 342.744 &  30.657 & 559.35 &   50.03 \\
    1542461401838152960 & 12:27:29.4 & +47:44:19.35 & 0.906 &   1.123 &  726.781 & 5128 &   0.461 &   0.273 &   2.29 &    1.36 \\
    6127569097490978944 & 12:32:38.4 & -48:58:46.68 & 0.834 &   0.761 & 1064.489 & 5028 &   0.268 &   0.238 &   4.19 &    3.72 \\
    3745090711928464512 & 13:28:22.2 & +15:50:07.71 & 1.423 &   4.948 &  381.228 & 4923 &   3.403 &   4.767 &   2.44 &    3.42 \\
    5864984550877297280 & 13:33:44.0 & -63:40:18.36 & 4.101 &   6.066 &  466.367 & 5480 &   2.721 &  30.046 &   3.57 &   39.47 \\
    5846761386917033600 & 14:20:27.4 & -68:26:19.33 & 2.496 &   7.522 &  334.953 & 5412 &   4.816 &  24.487 &   2.34 &   11.92 \\
    5785826933521556096 & 14:28:49.0 & -78:04:46.02 & 1.271 &   2.141 &  497.625 & 5982 &   1.025 &   0.850 &   1.64 &    1.36 \\
    4429810650412646528 & 15:43:19.6 & +06:39:15.87 & 1.050 &   1.776 &  621.893 & 5505 &   0.852 & \nodata &   2.65 & \nodata \\
    5819146877495757568 & 15:50:47.4 & -70:21:46.30 & 2.803 &   8.855 &  324.515 & 4856 &   5.806 &  17.370 &   2.57 &    7.69 \\
    5820369946745384832 & 15:51:00.8 & -68:49:32.34 & 2.354 &   9.470 &  280.023 & 4907 &   7.006 &   9.687 &   1.99 &    2.75 \\
    5931568402480842496 & 16:17:57.2 & -55:56:22.89 & 0.882 &   0.852 &  698.242 & 5328 &   0.309 &   0.249 &   1.36 &    1.10 \\
    5944159666018686848 & 16:27:46.6 & -43:56:32.63 & 2.494 &   9.253 &  285.196 & 5266 &   6.574 &  37.976 &   1.98 &   11.41 \\
    1410477980245374848 & 16:31:55.3 & +47:35:10.52 & 1.193 &   1.882 &  472.610 & 5936 &   0.872 &   0.476 &   1.19 &    0.65 \\
    4126243594591367168 & 16:53:45.5 & -21:51:49.28 & 2.187 &   5.615 &  505.349 & 5618 &   3.319 & \nodata &   5.55 & \nodata \\
    4549835637507855104 & 17:30:41.8 & +16:12:19.14 & 1.138 &   1.754 &  623.865 & 5750 &   0.803 &   0.516 &   2.52 &    1.62 \\
    4596507058545940096 & 17:54:04.6 & +28:57:49.06 & 2.446 &   7.710 &  299.051 & 5316 &   5.049 &  11.272 &   1.75 &    3.90 \\
    2153650709938721280 & 18:51:22.8 & +57:28:19.36 & 0.940 &   1.435 &  528.804 & 5456 &   0.654 &   0.398 &   1.25 &    0.76 \\
    4514619143398364800 & 19:06:08.2 & +18:05:23.58 & 1.685 &   3.364 &  475.779 & 5834 &   1.753 & \nodata &   2.45 & \nodata \\
    4306343947116038656 & 19:10:29.8 & +07:09:18.36 & 1.112 &   1.738 &  506.247 & 5800 &   0.802 & \nodata &   1.35 & \nodata \\
    4294229905988953856 & 19:32:11.7 & +05:22:58.40 & 2.299 &   7.053 &  322.304 & 5204 &   4.557 & \nodata &   1.98 & \nodata \\
    2025421609519820160 & 19:37:07.6 & +28:10:53.60 & 3.301 &  14.351 &  202.179 & 5026 &  11.038 & \nodata &   1.18 & \nodata \\
    2048262142478593280 & 19:38:46.2 & +35:53:15.35 & 0.971 &   1.262 &  556.966 & 5727 &   0.531 &   0.304 &   1.19 &    0.68 \\
    2087298894057921536 & 19:51:59.2 & +50:05:29.32 & 0.996 &   1.500 &  531.876 & 5647 &   0.679 &   0.387 &   1.32 &    0.75 \\
    2029905482299047040 & 20:11:06.6 & +30:38:54.25 & 1.126 &   1.814 &  617.427 & 5767 &   0.849 & \nodata &   2.59 & \nodata \\
    2062243493661150720 & 20:18:31.9 & +39:57:28.40 & 5.710 &   5.961 &  423.467 & 5866 &   2.246 &   4.677 &   2.21 &    4.60 \\
    2060993353245576448 & 20:20:52.2 & +37:48:18.97 & 5.035 &   4.544 &  706.794 & 5761 &   1.592 &  31.347 &   7.28 &  143.34 \\
    2054348862739979648 & 20:23:38.7 & +32:17:29.13 & 1.187 &   1.887 &  519.595 & 5912 &   0.878 &  29.823 &   1.59 &   54.18 \\
    2243723836359839872 & 20:23:48.9 & +63:33:43.03 & 0.908 &   1.271 &  579.668 & 5340 &   0.555 &   0.301 &   1.40 &    0.76 \\
    6425649173675248768 & 20:25:22.6 & -66:33:49.97 & 1.011 &   1.519 &  514.342 & 5743 &   0.687 &   0.616 &   1.21 &    1.09 \\
    4218379306035449216 & 20:35:27.0 & -04:49:38.79 & 1.018 &   1.457 &  568.020 & 5744 &   0.643 & \nodata &   1.53 & \nodata \\
    2066328011855804544 & 20:39:56.6 & +41:22:25.27 & 6.413 & 125.273 &  116.742 & 4073 & 204.228 &  11.785 &   4.21 &    0.24 \\
    1859677511539011456 & 20:51:40.8 & +31:39:32.67 & 1.408 &   2.137 &  570.741 & 5967 &   0.971 &   1.129 &   2.34 &    2.72 \\
    1844666360320230144 & 20:59:35.1 & +26:14:23.74 & 1.303 &   2.266 &  439.616 & 5949 &   1.102 &   0.493 &   1.21 &    0.54 \\
    1977530417708089728 & 21:37:23.6 & +47:04:53.23 & 1.048 &   1.627 &  539.573 & 5715 &   0.748 &   0.391 &   1.52 &    0.80 \\
    1987682247232185088 & 22:27:55.3 & +48:54:36.02 & 1.638 &   3.433 &  588.116 & 5485 &   1.833 &   8.539 &   4.83 &   22.50 \\
    1931538743969686912 & 23:05:51.3 & +42:32:43.33 & 1.116 &   1.755 &  611.487 & 5825 &   0.812 &   0.428 &   2.40 &    1.27 \\
    \enddata
    \tablecomments{(1 - 4) The mass ($M_1$), radius ($R_1$), radial velocity variation ($\DRV$),
      and temperature ($\Teff$) measured by \Gaia. (5) The theoretical minimum orbital
      period ($P^{\rm min}_{\rm orb}$) calculated from the \Gaia parameters. (6) The photometric
      period ($P^{\rm lc}_{\rm orb}$) folded from the \TESS light curve. (7) The mass function
      of the invisible components ($\massfunc_{\rm Gaia}$) calculated from $P^{\rm min}_{\rm orb}$. (8) The
      mass function of the invisible components ($\massfunc_{\rm lc}$) calculated from $P^{\rm lc}_{\rm orb}$.
      \label{table:2}}
  \end{deluxetable*}

  \bibliography{main}{}
  \bibliographystyle{aasjournal}

\end{CJK*}
\end{document}